\pdfoutput=1
\documentclass[letterpaper]{article}
\usepackage{aaai20}
\usepackage{times}
\usepackage{helvet}
\usepackage{courier}

\newcommand{\argmax}{\operatornamewithlimits{arg\,max}}
\usepackage{amsfonts}
\usepackage{siunitx}
\usepackage{multirow}
\usepackage{enumerate}
\usepackage{booktabs}
\usepackage{amssymb}
\usepackage{calc}
\usepackage{paralist}
\usepackage[labelfont=bf,textfont=bf,singlelinecheck=on]{subcaption}
\usepackage{tikz}
\usetikzlibrary{fit,positioning}
\usepackage{graphicx}
\usepackage[justification=justified]{caption}
\usepackage{diagbox}
\usepackage{color}
\usepackage{colortbl}
\usepackage{url}
\usepackage{mathtools}
\usepackage{algorithm}
\usepackage{algorithmic}
\usepackage[normalem]{ulem}
\usepackage{enumitem}
\usepackage{numprint}
\npthousandsep{,}
\npdecimalsign{.}
\usepackage[skip=0pt]{caption}
\usepackage{setspace}
\usepackage{color}
\usepackage{colortbl}
\usepackage{acronym}

\acrodef{SR}{Sequential Recommendation}
\acrodef{IDSR}{Intent-aware Diversified Sequential Recommendation}
\acrodef{IIM}{Implicit Intent Mining}
\acrodef{IDP}{Intent-aware Diversity Promoting}
\acrodef{GRU}{Gated Recurrent Unit}
\acrodef{MC}{Markov Chains}
\acrodef{RNN}{Recurrent Neural Network}
\acrodef{FPMC}{Factorizing Personalized Markov Chain}
\acrodef{HRNN}{Hierarchical Recurrent Neural Network}
\acrodef{NARM}{Neural Attentive Recommendation Machine}
\acrodef{MMR}{Maximal Marginal Relevance}
\acrodef{LTR}{Learning To Rank}
\acrodef{DPP}{Determinantal Point Process}
\acrodef{ILD}{Intra-List Distance}

\newcommand{\citet}[1]{\citeauthor{#1} \shortcite{#1}}
\newcommand{\citep}{\cite}

\newcommand{\noop}[1]{}

\newcommand{\chen}[1]{\textcolor{black}{#1}}

\newcommand{\duwen}{\phantom{$^\blacktriangle$}}

\title{Improving End-to-End Sequential Recommendations\\ with Intent-aware Diversification}
\author{Wanyu Chen$^{1,\,2}$\qquad Pengjie Ren$^{2}$
\qquad Fei Cai$^1$\qquad Maarten de Rijke$^2$\\
$^1$~National University of Defense Technology, Changsha, China\\ $^2$~University of Amsterdam, Amsterdam, The Netherlands\\{wanyuchen, caifei}@nudt.edu.cn, {p.ren, M.deRijke}@uva.nl
}

\begin{document}
\maketitle
\begin{abstract}
%
\acp{SR} that capture users' dynamic intents by modeling user sequential behaviors can recommend closely accurate products to users. 
Previous work on \acp{SR} is mostly focused on optimizing the recommendation accuracy, often ignoring the recommendation diversity, even though it is an important criterion for evaluating the recommendation performance.
Most existing methods for improving the diversity of recommendations are not ideally applicable for \acp{SR} because they assume that user intents are static and rely on post-processing the list of recommendations to promote diversity.
We consider both recommendation accuracy and diversity for \acp{SR} by proposing an end-to-end neural model, called \ac{IDSR}.
Specifically, we introduce an \ac{IIM} module into \acp{SR} to capture different user intents reflected in user behavior sequences.
Then, we design an \ac{IDP} loss to supervise the learning of the \ac{IIM} module and force the model to take recommendation diversity into consideration during training.
Extensive experiments on two benchmark datasets show that \ac{IDSR} significantly outperforms state-of-the-art methods in terms of recommendation diversity while yielding comparable or superior recommendation accuracy.
\end{abstract}

\section{Introduction}
\label{introduction}

Conventional recommendation methods often assume that user intents are static; they ignore the dynamic and evolving characteristics of user behavior.
\acfp{SR} have been introduced to address this issue; they aim to predict the next item(s) by modeling the sequence of a user's previous behavior~\cite{Quadrana:2018:SRS:3236632.3190616}. 

Early studies into \acp{SR} are mostly based on \ac{MC}~\cite{FPMC2010}.
Due to the unmanageable state space issue of \acp{MC}~\cite{2016session-based}, \ac{RNN} or Transformer based neural models have attracted a lot of attentions \cite{self-attentive18}.
A number of studies have investigated various factors that might influence \ac{SR} performances, e.g., personalization~\cite{HRNN2017}, repeat consumption \cite{ren-2019-repeatnet}, context~\cite{Rakkappan:2019:CSR:3308558.3313567}, etc.
However, these methods focus on improving recommendation accuracy only, which might have the risk of over-specialization, i.e., the recommended items are super homogeneous.

\begin{figure}[t]
 \centering
   \includegraphics[clip,trim=0mm 0mm 0mm 0mm,width=\columnwidth]{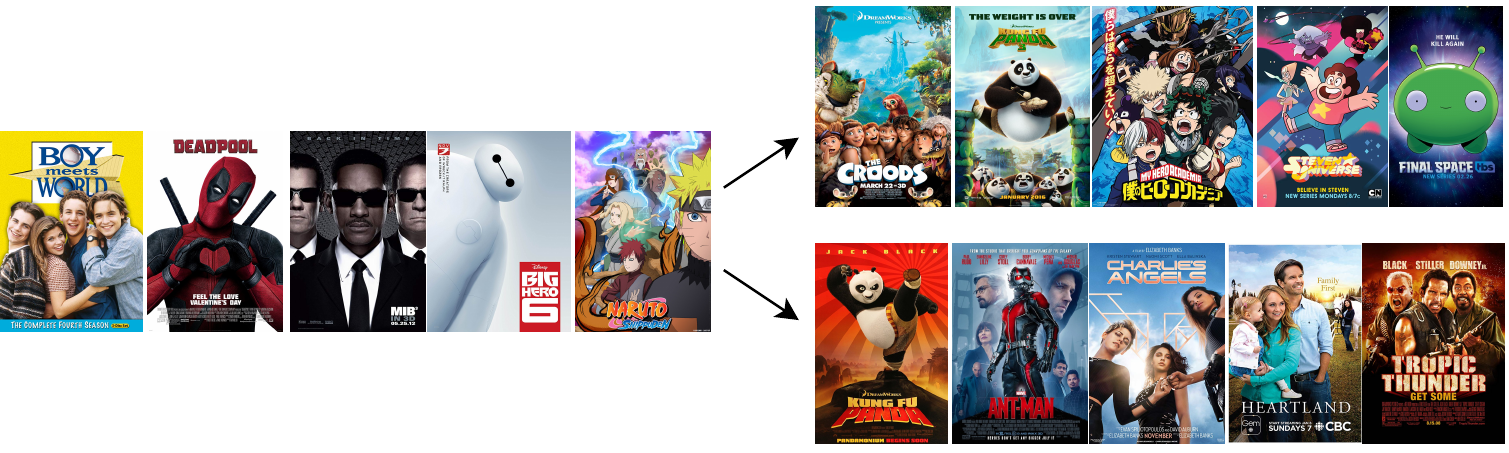}
   \caption{A example showing sequential recommendation with (bottom) and without (top) diversification.}
\label{example}
\end{figure}

This is problematic considering that users usually have multiple intents.
For example, as shown in Fig.~\ref{example}, although the user shows most interest in cartoon movies from her/his historic watching behaviors, s/he also watches family and action movies occasionally.
A better recommendation strategy should provide a diverse recommendation list to satisfy all these intents. In the case of Fig.~\ref{example}, we should recommend a list containing action, cartoon and family movies simultaneously instead of only cartoons.
Besides, sometimes, user intents are exploratory which means they do not have a specific intent in mind.
Thus homogeneous recommendation lists cannot satisfy such users and they easily get bored with the low diverse recommendation lists~\cite{Sha:2016:FRR}.

Diversification has been well studied in conventional recommendations~\cite{diversity-advances} as well as Web search~\cite{Abid2016}.
Current approaches for diversified recommendation mainly focus on how to re-rank the items based on a  certain diversity metric with general recommendation models.
However, they are not ideally applicable for \acp{SR} for two reasons.
First, some of them assume user intents are static and require that user intents are prepared beforehand, which is unrealistic in most \ac{SR} scenarios~\cite{Ashkan:2015:OGD,Chen:2018:FGM}.
Second, most of them belong to the post-processing paradigm and achieve recommendation accuracy and diversity in two separate steps, i.e., 1) scoring items and generating a candidate item set with a recommendation model; and 2) selecting a diverse recommendation list based on both the item scores and some implicit/explicit diversity metrics~\cite{diversity-advances,diversity-survey}.
Because the recommendation models are not aware of diversity during learning and it is hard to design ideal diversity strategies for different recommendation models, these methods are generally inferior and far from satisfactory.

To address the above issues, we take into account both recommendation accuracy and diversity, and propose an end-to-end neural model, namely \acf{IDSR}, for \acp{SR}.
Generally, \ac{IDSR} employs an \acf{IIM} module to automatically capture multiple latent user intents reflected in user behavior sequences, and directly generate accurate and diverse recommendation lists w.r.t the latent user intents.
In order to supervise the learning of the \ac{IIM} module and force the model to take recommendation diversity into consideration during training, we design an \acf{IDP} loss which evaluates recommendation accuracy and diversity based on the whole generated recommendation lists.
Specifically, a sequence encoder is first used to encode the user behaviors into representations.
Then, the \ac{IIM} module employs multiple attentions to mine user intents with each attention capturing a particular latent user intent.
Finally, an intent-aware recommendation decoder is used to generate a recommendation list by selecting one item at a time.
Especially, when selecting the next item, \ac{IDSR} also takes the already selected items as input so that it can track the satisfaction degree of each latent user intent. 
During training, we devise the \ac{IDP} loss to instruct \ac{IDSR} to learn to mine and track user intents.
All parameters are learned in an end-to-end back-propagation training paradigm within a unified framework.
We conduct extensive experiments on two benchmark datasets. 
The results show that \ac{IDSR} outperforms the state-of-the-art baselines on two publicly available datasets in terms of both accuracy metrics, i.e., Recall and MRR, and diversity metric, ILD.

Our contributions can be summarized as follows:
\begin{itemize}[nosep]
\item We propose an \acf{IDSR} method, which is the first end-to-end neural framework that considers diversification for \acp{SR}, to the best of our knowledge.
\item We devise an \acf{IIM} module to automatically mine latent user intents from user behaviors and an intent-aware recommendation decoder to generate diverse recommendation lists.
\item We present an \ac{IDP} loss to better supervise \ac{IDSR} in terms of recommendation accuracy and diversity.
\item We carry out extensive experiments and analyses on two benchmark datasets to verify the effectiveness of \ac{IDSR}.
\end{itemize}
\section{Related Work}
\label{related}

\begin{figure*}[t]
  \centering
   \includegraphics[clip,trim=0mm 0mm 0mm 0mm,width=\textwidth]{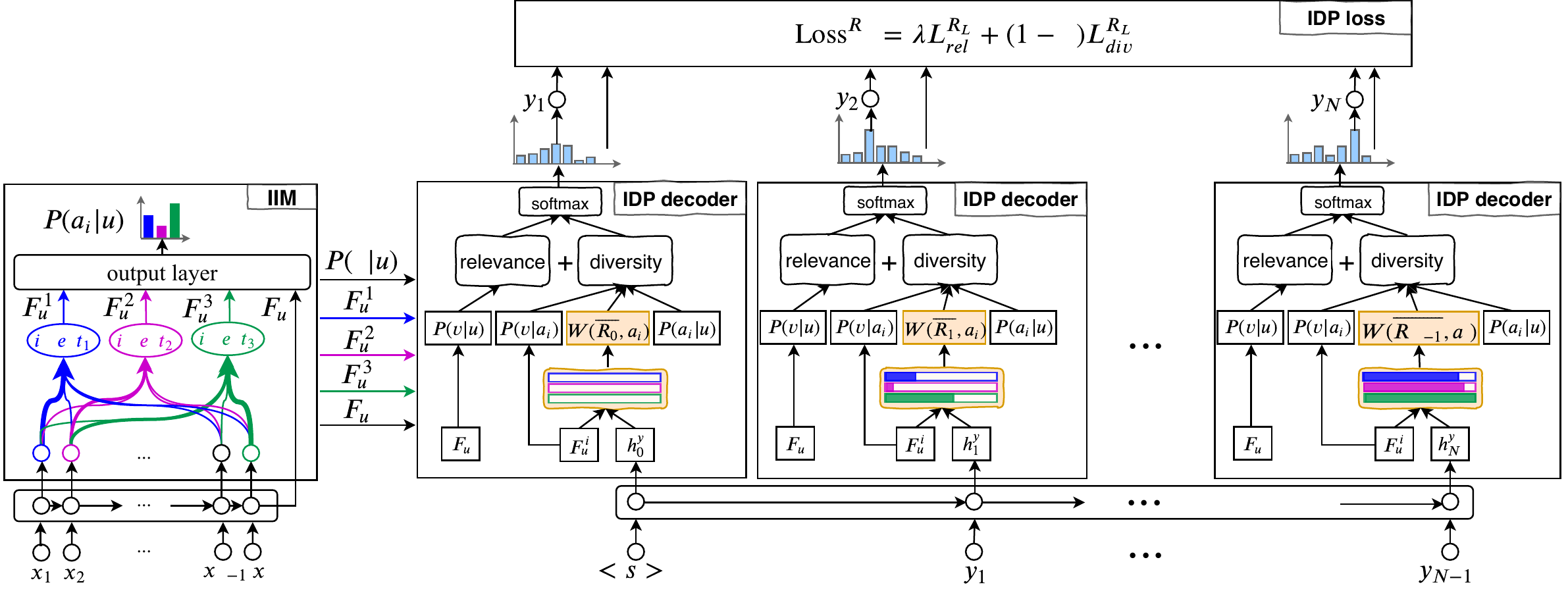}
   \caption{Overview of \ac{IDSR}. Blue, purple and green colors denote different user intents.}
\label{IDSR}
\end{figure*}

\subsection{Sequential recommendation}
Traditional methods are mainly based on \acp{MC}~\cite{Zimdars:2001}, which investigate how to extract sequential patterns to learn users' next preferences with probabilistic decision-tree models. 
Following this idea, \citet{Fusing2016} fuse similarity models with \acp{MC} for \acp{SR} to solve sparse recommendation problems. 
However, \ac{MC}-based methods only model local sequential patterns between adjacent interactions, which could not take the whole sequence into consideration.

Recently, \acp{RNN} have been devised to model variable-length sequential data. 
\citet{GRU2016} introduce an \ac{RNN}-based model for \acp{SR} that consists of \acp{GRU} and uses a session-parallel mini-batch training process.
\citet{HRNN2017} extend this idea and develop a hierarchical \ac{RNN} structure which takes users' profile into account by considering cross-session information.
Attention mechanisms have been applied to recommendation tasks to help models exploit users' preferences better~\cite{he2018nais}.
\citet{NARM2017} propose a neural attentive session-based recommendation machine that takes the last hidden state from the session-based \ac{RNN} as the sequential behavior, and uses the other hidden states of previous clicks for computing attention to capture users' current preferences in a given session. 
\citet{Xu:2019:RCN} propose a novel Recurrent Convolutional Neural Network model to capture both long-term as well as short-term dependencies for sequential recommendation.
Recently, a purely attention-based sequence-to-sequence model, Transformer, has achieved competitive performance on machine translation tasks~\cite{NIPS2017_7181}. 
\citet{self-attentive18} introduce Transformer into \acp{SR} by presenting a two-layer Transformer model to capture user’s sequential behaviors.
\citet{BERT2019} introduce a Bidirectional Encoder Representations from Transformers for sequential recommendation.

Although there are a number of studies for \acp{SR}, they only focus on accuracy of the recommendation list. 
None of the aforementioned studies has considered users' multiple intents and the diversification for \acp{SR}.

\subsection{Diversified recommendation}
Promoting the diversity for recommendation or search results has received increasing research attentions.
The most representative implicit approach is \ac{MMR}~\cite{MMR98}. 
\ac{MMR} represents relevance and diversity by independent metrics and uses the notion of marginal relevance to combine the two metrics with a tradeoff parameter.   
\citet{Sha:2016:FRR} introduce a submodular objective function to combine relevance, coverage of user’s interests, and the diversity between items.
\acf{LTR} has also been exploited to address diversification~\cite{Cheng:2017:LRA}. 
\citet{Cheng:2017:LRA} first label each user by a set of diverse as well as relevant items with a heuristic method and then propose a diversified collaborative filtering algorithm to learn to optimize the performance of accuracy and diversity for recommendation.
The main problem is that they all need diversified ranking lists as ground-truth for learning~\cite{diversity-advances}, which is usually unavailable in recommendations.
Recently, \citet{Chen:2018:FGM} propose to improve recommendation diversification through \ac{DPP} \cite{Kulesza:2012:DPP} with a greedy maximum a posterior inference algorithm.

All above methods achieve recommendation accuracy and diversity in two separate processes, i.e., training an offline recommendation model to score items in terms of accuracy and then re-ranking items by taking diversity into account. 
In addition, they are not suitable for \acp{SR} where users' sequential behaviors need to be considered.


\section{\acl{IDSR}}
\label{approach}

\subsection{Overview}
Given a user $u$ and her/his behavior sequence $S_u$=$\{x_1,x_2,\ldots,x_T\}$ ($x_i$ is the item that $u$ interacts with, e.g., watched movies), our goal is to provide $u$ with a recommendation list $R_L$ for predicting her/his next interaction, of which the items are expected to be both relevant and diverse.

Different from existing \ac{SR} methods, we assume there are $M$ latent intents behind each behavior sequence, i.e., $A$=$\{a_1,\ldots,a_M\}$.
Then, we seek to generate a recommendation list $R_L$ by maximizing the satisfactory degree of all intents.
\begin{equation}
\label{equation1}
 \mbox{}\!\!P(R_L\mid u,S_u) = \sum^M_{i=1} P(a_i\mid u)P(R_L\mid a_i,u,S_u),
\end{equation}
where $P(a_i\mid u)$ denotes the importance of the intent $a_i$ to user $u$. 
$P(R_L\mid a_i,u,S_u)$ is the satisfactory probability of $R_L$ to $a_i$.

It is hard to directly optimize $P(R_L\mid u,S_u)$ due to the huge search space.
Therefore, we propose to generate $R_L$ greedily, i.e., selecting one item at a time with the maximum score $S(v)$.
\begin{equation}
\label{equation2}
   \mbox{}\!\!v_{t}\gets \argmax_{v\in V\backslash R_{t-1}}S(v),
\end{equation}
where $v_{t}$ is the item to be selected at step $t$; $V$ is the set of all items; $R_{t-1}$ is the generated recommendation list until step ${t-1}$;
$V\backslash R_{t-1}$ guarantees that the selected item is different from previous generated recommendations in $R_{t-1}$ at step $t$.
$S(v)$ returns the score of item $v$ by
\begin{equation}
\label{equation3}
\begin{aligned}
S(v)\gets&\lambda P(v\mid u,S_u)\\
&+(1-\lambda)\sum^M_{i=1} P(a_i\mid u)P(v\mid a_i)W(\overline{R_{t-1}},a_i).
\end{aligned}
\end{equation}
Generally, it is a combination of the relevance score and diversification score balanced by a hyper-parameter $\lambda$.
$P(v\mid u,S_u)$ is the relevance score reflecting the interest of $u$ on v.
$P(a_i\mid u)$ is the probability of $u$ with the intent $a_i$.
$P(v\mid a_i)$ is the satisfactory degree of $v$ to $a_i$.
$W(\overline{R_{t-1}},a_i)$ denotes the likelihood that the already generated recommendation list $R_{t-1}$ cannot satisfy $a_i$.

We propose an end-to-end \ac{IDSR} model to directly generate a diversified recommendation list upon Eq.~\ref{equation3}.
As shown in Fig.~\ref{IDSR}, \ac{IDSR} consists of three modules: a Sequence encoder, an \acf{IIM} module and an \acf{IDP} decoder.
The sequence encoder encodes users' sequential behaviors into latent representations.
Then, the \ac{IIM} module is used to capture users' multiple latent intents reflected in the sequential behaviors.
Finally, the \ac{IDP} decoder is employed to generate a recommendation list w.r.t. Eq.~\ref{equation3}.
We devise an \ac{IDP} loss to train \ac{IDSR} which evaluates the whole recommendation list in terms of both recommendation accuracy and diversity.
Note that there is no re-ranking involved in \ac{IDSR} and both recommendation accuracy and diversity are jointly learned in an end-to-end way.
Next, we introduce the separate modules.
\subsection{Sequence encoder}
Two commonly used technologies for sequence modeling are \ac{GRU} and Transformer. 
We use both in our experiments (see \S\ref{results}) and find that \ac{IDSR} shows better performance with \ac{GRU}. 
Thus here we use a \ac{GRU} to encode $S_u$.
\begin{equation}
\label{equation4}
\begin{aligned}
z_t&=\sigma(W_z[\mathbf{x_t},h_{t-1}])\\
r_t&=\sigma(W_r[\mathbf{x_t},h_{t-1}])\\
\hat{h_t}&=\tanh(W_h[\mathbf{x_t},r_t\odot h_{t-1}])\\
h_t&=(1-z_t)\otimes h_{t-1}+z_t\odot \hat{h_t},
\end{aligned}
\end{equation}
where $\mathbf{x_t}$ denotes the embedding of item $x_t$; $W_z$, $W_r$ and $W_h$ are weight parameters; $\sigma$ denotes the sigmoid function. 
The inputs of the encoder is the behavior sequence $S_u$=$\{x_1,x_2,\ldots,x_T\}$ and the outputs are hidden representations $\{h_1,h_2,\ldots,h_T\}$, where $h_i\in\mathbb{R}^{d_e}$. 
We stack those representations into matrix $H_S\in\mathbb{R}^{T\times d_e}$.
Generally, we consider the last representation $h_T$ as a summary of the whole sequence~\cite{NARM2017}.
Thus we set the global preference $F_u=h_T$.

\subsection{\ac{IIM} module}
The \ac{IIM} module is to evaluate the $P(a_i\mid u)$ in Eq.~\ref{equation3}.
Intuitively, a user's multiple intents can be reflected by different interactions in the sequential behaviors. 
Some interactions are more representative for a particular intent than the other interactions, e.g., the last two behaviors in Fig.~\ref{example} reflects the intent of watching cartoon movies.
Motivated by this, we fuse a multi-intent attention mechanism with each attention to capture one particular intent.
Specifically, \ac{IIM} first projects $H_S$ and $F_u$ into $M$ spaces w.r.t. the latent intents respectively. 
Then, $M$ attention functions are employed in parallel to produce the user's intent-specific representations $\{F^1_u,F^2_u,\ldots,F^M_u\}$.
\begin{equation}
\label{equation5}
F^i_u=\text{Attention}(F_uW^Q_i, H_SW^K_i,H_SW^V_i)
\end{equation}
where the projection matrices for intent $i$, i.e., $W^Q_i\in\mathbb{R}^{d_e\times d}$, $W^K_i\in\mathbb{R}^{d_e\times d}$ and $W^V_i\in\mathbb{R}^{d_e\times d}$, are learnable parameters.
We use the scaled dot-product attention in this work~\cite{ATTention2017}.
After that, the importance of each intent, i.e., $P(a_i\mid u)$ in Eq.~\ref{equation3}, can be calculated by
\begin{equation}
\label{equation6}
\begin{aligned}
P(a_i\mid u)&=\frac{\text{exp}({F_u}^\mathrm{T}F^i_u)}{\sum^M_{j=1}\text{exp}({F_u}^\mathrm{T}F^j_u)}.
\end{aligned}
\end{equation}

\subsection{\ac{IDP} decoder}
The \ac{IDP} decoder is to generate $R_L$ based on the intents mined with the \ac{IIM} module.
To begin with, we model the relevance score of $v$ to user $u$ (i.e., $P(v\mid u,S_u)$ in Eq.~\ref{equation3}) as follows:
\begin{equation}
\label{equation7}
P(v\mid u,S_u)=\frac{\text{exp}(\mathbf{v}^\mathrm{T}F_u)}{\sum^{|V|}_{j=1}\text{exp}(\mathbf{v_j}^\mathrm{T}F_u)}.
\end{equation}

Similarly, we model the relevance of $v$ to intent $a_i$ (i.e., $P(v\mid a_i)$ in Eq.~\ref{equation3}) as follows:
\begin{equation}
\label{equation8}
P(v\mid a_i)=\frac{\text{exp}(\mathbf{v}^\mathrm{T}F^i_u)}{\sum^{|V|}_{n=1}\text{exp}(\mathbf{v_n}^\mathrm{T}F^i_u)}.
\end{equation}

To track the already selected items to date, we use another \ac{GRU} to encode $R_{t-1}$=$\{y_1,y_2,\ldots,y_{t-1}\}$ into $\{h^y_1,h^y_2,\ldots,h^y_{t-1}\}$.
Then we estimate the satisfactory of $R_{t-1}$ to each intent (i.e., $W(\overline{R_{t-1}},a_i)$ in Eq.~\ref{equation3}) by calculating the matching between $h^y_{t-1}$ and $F^i_u$ as follows:
\begin{equation}
\label{equation9}
\begin{aligned}
W(\overline{R_{t-1}},a_i)&=\frac{\text{exp}(w^i_{t-1})}{\sum^M_{j=1}\text{exp}(w^j_t)},\\
w^i_{t-1}&=-h^y_{t-1}W_yF^i_u.
\end{aligned}
\end{equation}

Finally, we can calculate the score $S(v)$ of each item (Eq.~\ref{equation3}), select the item with highest probability, and append it to the recommendation list. 

\subsection{\ac{IDP} loss}
Since our goal is to generate a recommendation list which is both relevant and diverse, we design the \ac{IDP} loss function to evaluate the whole generated list $R_L$:
\begin{equation}
\label{equation10}
\text{Loss}^{R_L}=\lambda L^{R_L}_{rel}+(1-\lambda)L^{R_L}_{div},
\end{equation}
where $\lambda$ is a trade-off parameter, which is the same one as that in Eq 3, 
\chen{as they both control the contributions of accuracy and diversification. When $\lambda$ increases, we consider more about accuracy in both of \ac{IDP} decoder and \ac{IDP} loss.}
Given the output recommendation list from \ac{IDSR} $R_L=\{y_1,y_2,\ldots,y_N\}$ and the ground-truth item $y^*$ (i.e., the next consumed item), $L^{R_L}_{rel}$ is defined as follows:
\begin{equation}
\label{equation11}
L^{R_L}_{rel}=\left\{
\begin{array}{ll}
-\frac{1}{N}\displaystyle\sum^N_{t=1}t\cdot\mathbb{I}(y_t,y^*)\log \mathbf{p}^*_t
, & y^*\in R_L\\
-\frac{1}{N}\displaystyle\sum^N_{t=1}\log\mathbf{p}^*_t, & y^*\not\in R_L,
\end{array}
\right.
\end{equation}
where $\mathbb{I}(y_i,y_{t})$ is an indicator function that equals 1 if $y_i=y_{t}$ and 0 otherwise.
$\mathbf{p}^*_t$ denotes the probability of the ground-truth item $y^*$ at $t$-th step when generating the recommendation list $R_L$.
$L^{R_L}_{rel}$ encourages the positive recommendation list which contains the ground-truth item $y^*$ and punishes the negative ones otherwise.
Note that $L^{R_L}_{rel}$ also takes the ranking position of the ground-truth item $y^*$ into consideration by weighting the loss with the position $t$.

To promote diversity, inspired by the result diversification in Web search~\cite{Agrawal:2009:DSR}, we define $L^{R_L}_{div}$ as the probability of each intent $a_i$ having at least one relevant item in $R_L$
\begin{equation}
\label{equation12}
\mbox{}\!\!L^{R_L}_{div}= -\sum^M_{i=1}P(a_i\mid u)\left(1-\prod_{v\in R_L}(1-P(v\mid a_i))\right),
\nonumber
\end{equation}
where $P(a_i\mid u)$ and $P(v\mid a_j)$ are defined in Eq.~\ref{equation6} and Eq.~\ref{equation8} respectively.

All parameters of \ac{IDSR} as well as the item embeddings can be learned in an end-to-end back-propagation training paradigm.


\section{Experimental Setup}
\label{ex-setup}

We seek to answer the following research questions:
\begin{enumerate}[label=(\textbf{RQ\arabic*}),leftmargin=*,nosep]
\item What is the performance of \ac{IDSR} compared with state-of-the-art baselines in terms of accuracy?
\item Does \ac{IDSR} outperform state-of-the-art baselines in terms of diversity?
\item What is the impact of different sequence encoders (i.e., \ac{GRU}, Transformer) on \ac{IDSR}?
\item How does the trade-off parameter $\lambda$ affect the performance of \ac{IDSR}?
\item How does \ac{IDSR} influence the recommendation results?
\end{enumerate}

\begin{table}[!t]
\centering
\caption{Dataset statistics.}
\label{dataset}
\begin{tabular}{lrr}
\toprule
    Dataset& ML100K & ML1M\\
    \midrule
Number of users & 943 & 6,040 \\
Number of items & 1,682 & 3,706 \\
Number of interactions & 100,000 &1,000,209\\
Number of item genres & 19 & 18 \\
Avg.\ number of genres per item & 1.7 & 1.6 \\
\bottomrule
\end{tabular}
\vspace*{-.5\baselineskip}
\end{table}

\subsection{Datasets}
To answer our research questions, we use two publicly available datasets for the experiments.
Table~\ref{dataset} lists the statistics of the two datasets.
\begin{itemize}[nosep, align=left, leftmargin=*]
    \item  \textbf{ML100K}\footnote{\url{https://grouplens.org/datasets/movielens/}} is collected from the MovieLens web site. It contains 100,000 ratings from 943 users on 1,682 movies.
    \item  \textbf{ML1M}\footnotemark[\value{footnote}] is a larger and sparser version of ML100K, which contains 4,607,047 ratings for movies~\cite{RSsurvey2017}. 
\end{itemize}

We do not use the datasets as in \cite{HRNN2017} because there are only item ids.
We cannot conduct diversity evaluation nor case study.
We preprocess ML100K and ML1M for \ac{SR} experiments with the following steps. 
First, we filter out users who have less than 5 interactions and the movies that are rated less than 5 times. 
Then, we sort the rated movies according to the ``{timestamp}'' field to get a sequence of behaviors for each user.
Finally, we prepare each data sample by regarding the former 9 behaviors as input and the next behavior as output.
For evaluation, we randomly divide the datasets into training ({70\%}), validation ({10\%}) and test (20\%) sets.
We make sure that all movies in the test set have been rated by at least one user in the training set and the test set contains the most recent behaviors which happened later than those in the training and validation sets.


\subsection{Comparison methods}
We select several traditional recommendation methods as well as recent state-of-the-art neural \ac{SR} methods as baselines.
\begin{itemize}[nosep, align=left, leftmargin=*]
\item \textbf{POP}: POP ranks items based on the number of interactions, which is a non-personalized approach~\cite{RS-CF2005}.
\item \textbf{FPMC}: FPMC is a hybrid model that combines \acp{MC} and collaborative filtering for \acp{SR}~\cite{FPMC2010}.
\item \textbf{GRU4Rec}: GRU4Rec is an \ac{RNN}-based model for \acp{SR}. GRU4Rec utilizes session-parallel mini-batches as well as a ranking-based loss function in the training process~\cite{GRU2016}.
\item \textbf{HRNN}: HRNN is a hierarchical \ac{RNN} for \acp{SR} based on GRU4Rec. It adopts a session-level \ac{RNN} and a user-level \ac{RNN} to model users' short-term and long-term preferences~\cite{HRNN2017}.
\end{itemize}

Because there is no previous work specific for diversified \acp{SR}, we construct three baselines \textbf{FPMC+MMR}, \textbf{GRU4Rec+MMR} and \textbf{HRNN+MMR} ourselves.
Specifically, we first get the relevance scores $S(v)$ for each item with FPMC, GRU4Rec or HRNN.
Then, we rerank the items using the \ac{MMR} criteria
\begin{equation}
\label{equation14}
\textstyle
\mbox{}\!\!v\gets\argmax_{v_i\in R_c\backslash R_{L}}\theta S(v_i)+(1-\theta)\min_{v_k\in R_{L}}d_{ki},
\nonumber
\end{equation}
where $R_c$ is a candidate item set and $\theta\in[0,1]$ is a trade-off parameter to balance the relevance and the minimal dissimilarity $d_{ki}$ between item $v_k$ and item $v_i$. 
\ac{MMR} first initializes $R_L=\emptyset$ and then iteratively selects the item into $R_L$, until $|R_L|=N$.
When $\theta$=0, MMR returns diversified recommendations without considering relevance; when $\theta$=1, it returns the same results as the original baseline models.
We cannot choose the best $\theta$ relying solely on accuracy or diversity.
To balance both, we set $\theta=0.5$ in our experiments.

In addition, we consider two variants of our \ac{IDSR} model:
\begin{itemize}[nosep, align=left, leftmargin=*]
\item \textbf{IDSR}$_{\mathit{trans}}$: We use transformer to encoder users' behavior sequences in \ac{IDSR}.
\item \textbf{IDSR}$_{\mathit{GRU}}$: We use \ac{GRU} to encoder users' behavior sequences in \ac{IDSR}.
\end{itemize}

\subsection{Evaluation metrics}
For accuracy evaluation, we use Recall and MRR~\cite{NARM2017,STAMP2018}; For diversity evaluation, we use \ac{ILD}~\cite{Zhang:2008:AMI}:
\begin{itemize}[nosep, align=left, leftmargin=*]
\item  \textbf{Recall}: A primary metric which is used to evaluate the recall of the recommender system, i.e., whether the test item is contained in the recommendation list.
\item  \textbf{MRR}: A metric measures the ranking accuracy of the recommender system, i.e., whether the test item is ranked at the top of the list.
\item  \textbf{ILD}: A commonly used metric which measures the diversity of a recommendation list as the average distance between pairs of recommended items. ILD is defined as:
\begin{equation}
\label{equation15}
\textstyle
\text{ILD}=\frac{2}{|R_L|(|R_L|-1)}\sum_{(i,j)\in R_L}d_{ij}.
\end{equation}
We calculate the dissimilarity $d_{ij}$ between two movies based on Euclidean distance between the item genre vectors of movies~\cite{Ashkan:2015:OGD}.
\end{itemize}

\subsection{Implementation details}
We set the item embedding size and \ac{GRU} hidden state sizes to 100.
We use dropout with drop ratio $p = 0.1$.
We initialize model parameters randomly using the Xavier method~\cite{pmlr-v9-glorot10a}. 
We optimize the model using Adam~\cite{Adam2014} with the initial learning rate $\alpha = 0.001$, two momentum parameters $\beta1 = 0.9$ and $\beta2 = 0.999$, and $\epsilon = 10^{-8}$. The mini-batch size is set to 128.
We test the model performance on the validation set for every epoch. 
The code used to run the experiments is available online\footnote{\url{https: //url.suppressed.for.anonymity}}.
\section{Results and Analysis}
\label{results}
\begin{table*}[!ht]
\captionsetup{justification=justified}
   \centering
   \caption{Performance of recommendation models. The results produced by the best baseline and the best performer in each column are underlined and boldfaced, respectively. Statistical significance of pairwise differences of IDSR$_{\mathit{GRU}}$ vs.\ the best baseline is determined by a paired $t$-test ($^\blacktriangle$ for {p-value} $\leq$ .01, or $^\vartriangle$ for {p-value} $\leq$ .05).}
 \label{performance}
 \setlength{\tabcolsep}{0.85mm}{
 \begin{tabular}{lccccccccccccccccc}
 \toprule
  &\multicolumn{8}{c}{ML100K} && \multicolumn{8}{c}{ML1M} \\
  \cmidrule{2-9}\cmidrule{11-18}
 & \multicolumn{2}{c}{Recall} && \multicolumn{2}{c}{MRR}&& \multicolumn{2}{c}{ILD} && \multicolumn{2}{c}{Recall} && \multicolumn{2}{c}{MRR} && \multicolumn{2}{c}{ILD} \\
 \cmidrule{2-3}\cmidrule{5-6}\cmidrule{8-9}\cmidrule{11-12}\cmidrule{14-15}\cmidrule{17-18}
 Model & @10 & @20 && @10 & @20 && @10 & @20&& @10 & @20&& @10 & @20&& @10 & @20 \\
 \midrule
 POP & .0397\duwen & .0804\duwen && .0131\duwen & .0166\duwen && 3.330\duwen & 3.401\duwen&& .0578\duwen & .1121\duwen&& .0181\duwen & .0277\duwen && 2.518\duwen &2.681\duwen \\
 FPMC & .0411\duwen & .1011\duwen && .0155\duwen & .0207\duwen && 3.011\duwen & 3.071\duwen&& .0731\duwen & .1344\duwen&& .0201\duwen & .0447\duwen && 2.001\duwen & 2.017\duwen \\
 GRU4Rec &  .1082\duwen & .1787\duwen && .0321\duwen & .0371\duwen && 3.121\duwen & 3.144\duwen&& .1101\duwen & .1956\duwen&& .0399\duwen & .0501\duwen && 2.006\duwen & 2.212\duwen \\
 HRNN & \underline{.1106}\duwen & \underline{.1897}\duwen && \underline{.0362}\duwen & \underline{.0402}\duwen && 3.148\duwen & 3.164\duwen&& .\underline{1171}\duwen & \underline{.2143}\duwen&& \underline{.0418}\duwen & \underline{.0545}\duwen && 2.045\duwen & 2.224\duwen \\
 \midrule
 FPMC+MMR & .0397\duwen & .0971\duwen && .0121\duwen & .0197\duwen && 3.374\duwen & 3.415\duwen&& .0701\duwen & .1332\duwen&& .0199\duwen & .0431\duwen && 2.865\duwen & 2.932\duwen \\
 GRU4Rec+MMR &  .1037\duwen & .1711\duwen && .0311\duwen & .0361\duwen && 3.493\duwen & 3.506\duwen&& .1043\duwen & .1901\duwen&& .0383\duwen & .0489\duwen && 2.879\duwen & 2.985\duwen \\
 HRNN+MMR & .1094\duwen & .1857\duwen && .0351\duwen & .0396\duwen && \underline{3.515}\duwen & \underline{3.525}\duwen&& .1141\duwen & .2101\duwen&& .0411\duwen & .0532\duwen && \underline{2.911}\duwen & \underline{3.000}\duwen \\
 \midrule
IDSR$_{\mathit{GRU}}$& \bf{.1152}$^\blacktriangle$ & \bf{.2018}$^\blacktriangle$ && \bf{.0382}$^\blacktriangle$ & \bf{.0429}$^\blacktriangle$ && \bf{3.775}$^\blacktriangle$ & \bf{3.806}$^\blacktriangle$&& \bf{.1203}$^\vartriangle$  & \bf{.2212}$^\vartriangle$ && \bf{.0431}$^\vartriangle$  & \bf{.0574}$^\vartriangle$  && \bf{3.227}$^\blacktriangle$ & \bf{3.354}$^\blacktriangle$ \\
 \bottomrule
 \end{tabular}}
 \end{table*}

\subsection{Performance in terms of accuracy}
To answer RQ1, we examine the performance of \ac{IDSR} and the baseline models in terms of Recall and MRR; see Table~\ref{performance}. 

First, we can see that \ac{IDSR} outperforms the traditional methods, i.e., POP and FPMC, as well as the neural-based methods, i.e., GRU4Rec and HRNN in terms of Recall and MRR. 
When the size of recommendation list changes from 10 to 20, the improvements of \ac{IDSR} over the best baseline HRNN get increased. 
In detail, the improvements are 4.16\% and 6.37\% in terms of Recall@10 and Recall@20; 5.52\% and 6.72\% in terms of MRR@10 and MRR @20 on ML100K.  
Besides, the improvements of \ac{IDSR} over HRNN in terms of MRR are larger than that of Recall. 
For instance, on the ML1M dataset, the improvements of \ac{IDSR} over HRNN are 5.32\% in terms of MRR@20 while 3.22\% in terms of Recall@20. 
This shows that our model can not only boost the number of relevant items but the ranking of relevant items.
It may be due to the fact that our \ac{IIM} module can capture a user's multiple intents and their importances in his current preferences. Thus at the first steps in \ac{IDP} decoder, \ac{IDSR} can give high probabilities to relevant items.

Second,  we note that after re-ranking with \ac{MMR}, the performances of the baseline models drop a little bit in terms of both Recall and MRR.
This indicates that although post-processing with \ac{MMR} can improve the diversity of recommendation list, it might hurt the accuracy.
Because most of the candidate items generated by the baseline models have similar genres, the diversity score for the relevant item may be much lower than irrelevant ones. This might lead to a situation that the irrelevant items have higher final scores than the relevant item, which results in a worse performance.

\subsection{Performance in terms of diversity}
To answer RQ2, we report the ILD scores on both datasets in Table~\ref{performance}.
We can see that \ac{IDSR} consistently outperforms all baselines. 
The improvements of \ac{IDSR} over HRNN are 19.90\% and 20.29\% in terms of ILD@10 and ILD@20 on ML100K dataset, 57.76\% and 50.85\% on ML1M dataset.
Compared with the post-processing baselines, our model outperforms the best baseline model, i.e., HRNN+MRR, by 7.39\% and 7.97\% in terms of ILD@10 and ILD@20 on ML100K dataset and 10.83\% and 11.81\% on ML1M dataset.
Significant improvements of \ac{IDSR} against the best performing baseline are observed with a {paired $t$-test}.
This proves that our model is competitive in improving the diversity for sequential recommendation than post-processing methods.
It may be because that the effectiveness of post-processing methods, e.g., MMR, can be impacted by the baseline models.
When the candidate items generated by the sequential recommendation baselines are of similar genres, the performance of MMR method is limited to some extent.

\subsection{Performance with different encoders}
To answer RQ3, we test the performance of \ac{IDSR} with different sequence encoders, i.e, GRU or Transformer.  Table~\ref{encoder} shows the results. 

\begin{table}[!h]
\captionsetup{justification=justified}
  \centering
  \caption{Performance of \ac{IDSR} model with different sequence encoder (GRU or Transformer). The results produced by the best performer in each row are boldfaced.}
\label{encoder}
\begin{tabular}{llcc}
\toprule
Dataset&Metric&{IDSR}$_{\mathit{GRU}}$ &{IDSR}$_{\mathit{Trans}}$\\
\midrule
\multirow{3}{*}{ML100K}& Recall@20& \bf{\phantom{3}.2018} & \phantom{3}.1819\\
&MRR@20 & \bf{\phantom{3}.0429} & \phantom{3}.0385\\
&ILD@20 & \bf{3.806\phantom{3}}& 3.635\phantom{3}\\
\midrule
\multirow{3}{*}{ML1M}& Recall@20& \bf{\phantom{3}.2212} & \phantom{3}.2093\\
&MRR@20 & \bf{\phantom{3}.0574} & \phantom{3}.0529\\
&ILD@20 & \bf{3.354\phantom{3}} & 3.313\phantom{3}\\
\bottomrule
\end{tabular}
\end{table}

\begin{figure*}
        \begin{subfigure}[t]{0.32\textwidth}
                \includegraphics[width=\textwidth]{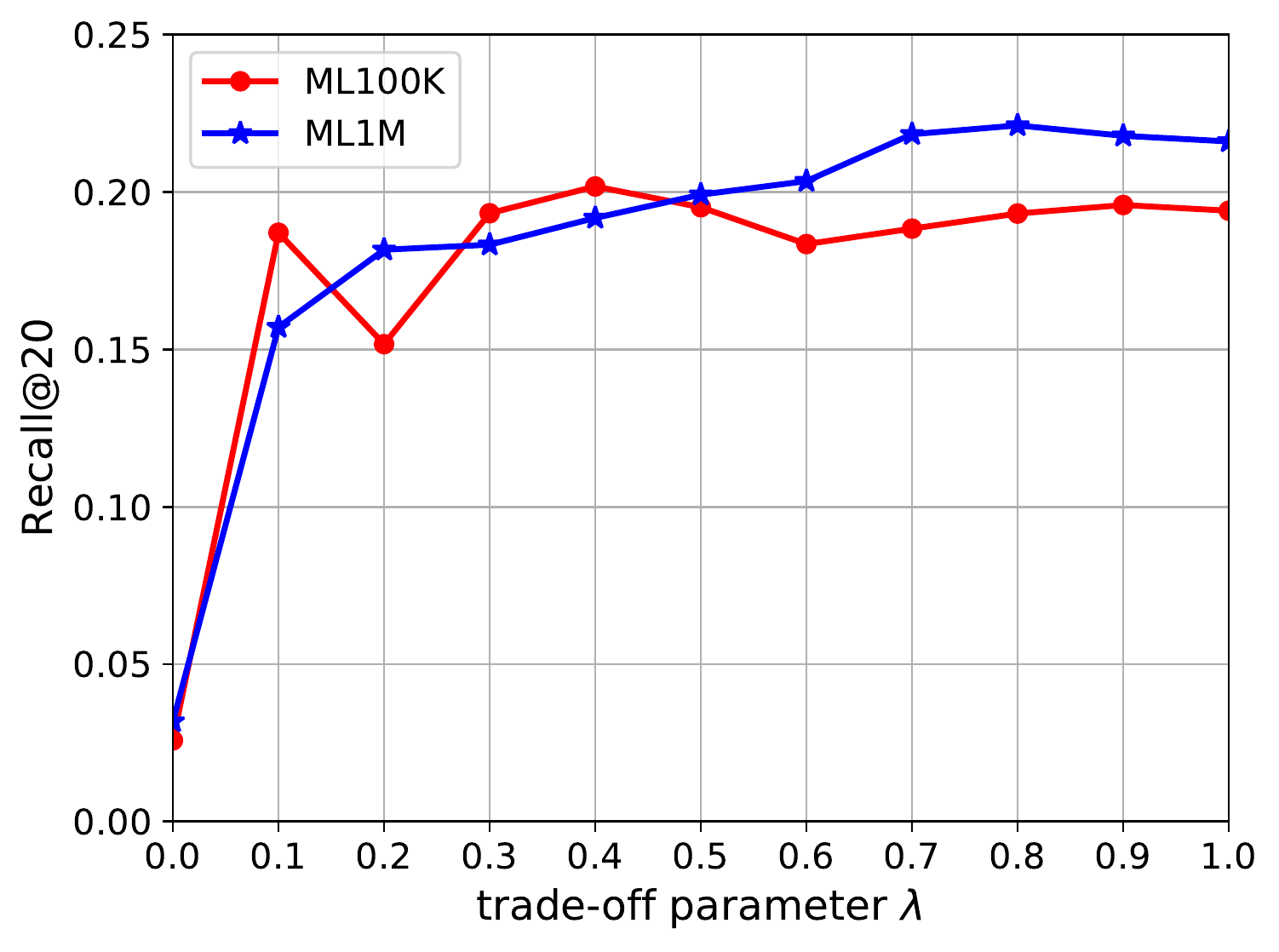}
                \caption{Performance in terms of Recall@20.}
                \label{recall}
        \end{subfigure}
        \begin{subfigure}[t]{0.32\textwidth}
                \includegraphics[width=\textwidth]{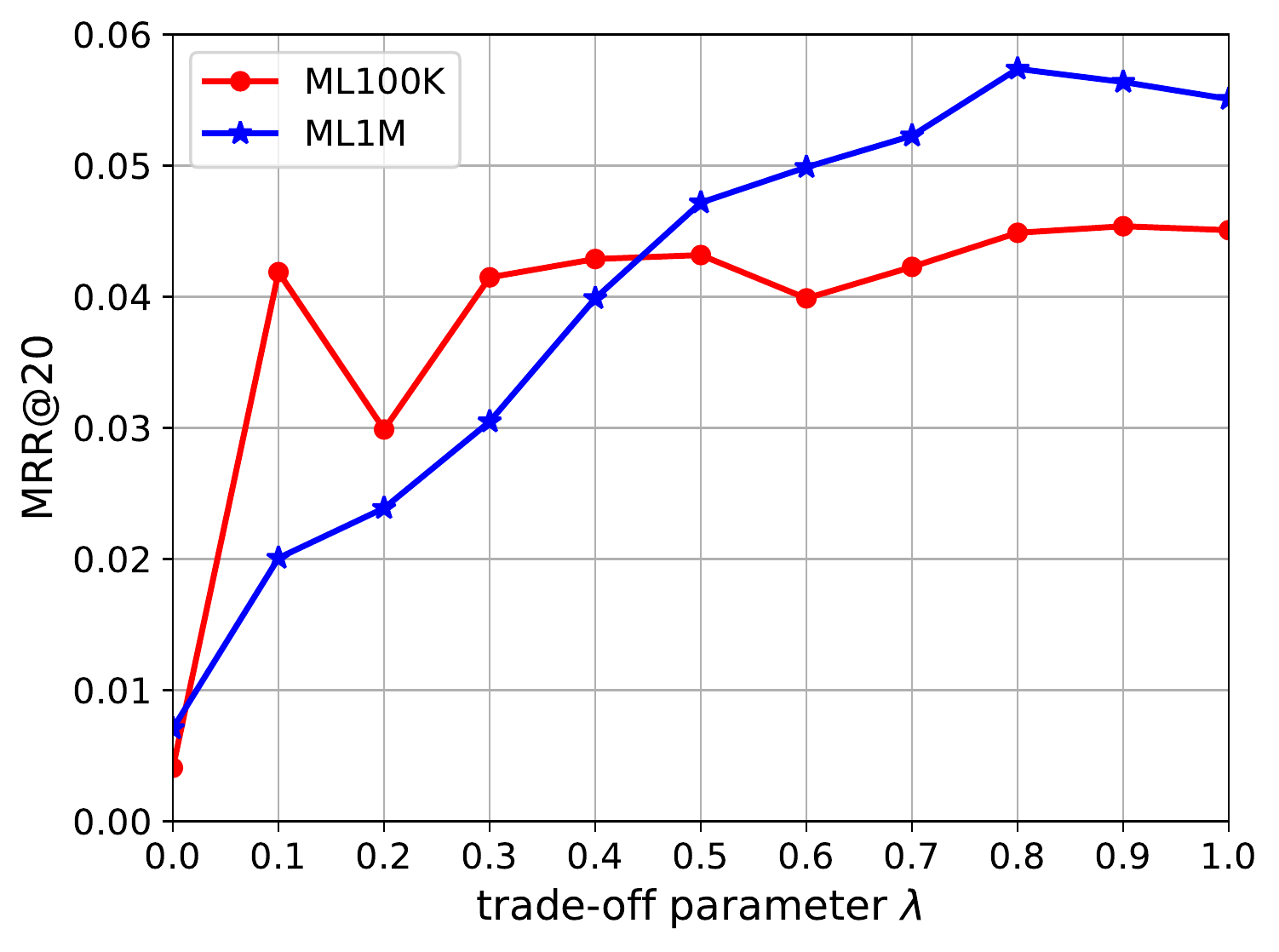}
                \caption{ Performance in terms of MRR@20.}
                \label{mrr}
        \end{subfigure}
        \begin{subfigure}[t]{0.32\textwidth}
                \includegraphics[width=\textwidth]{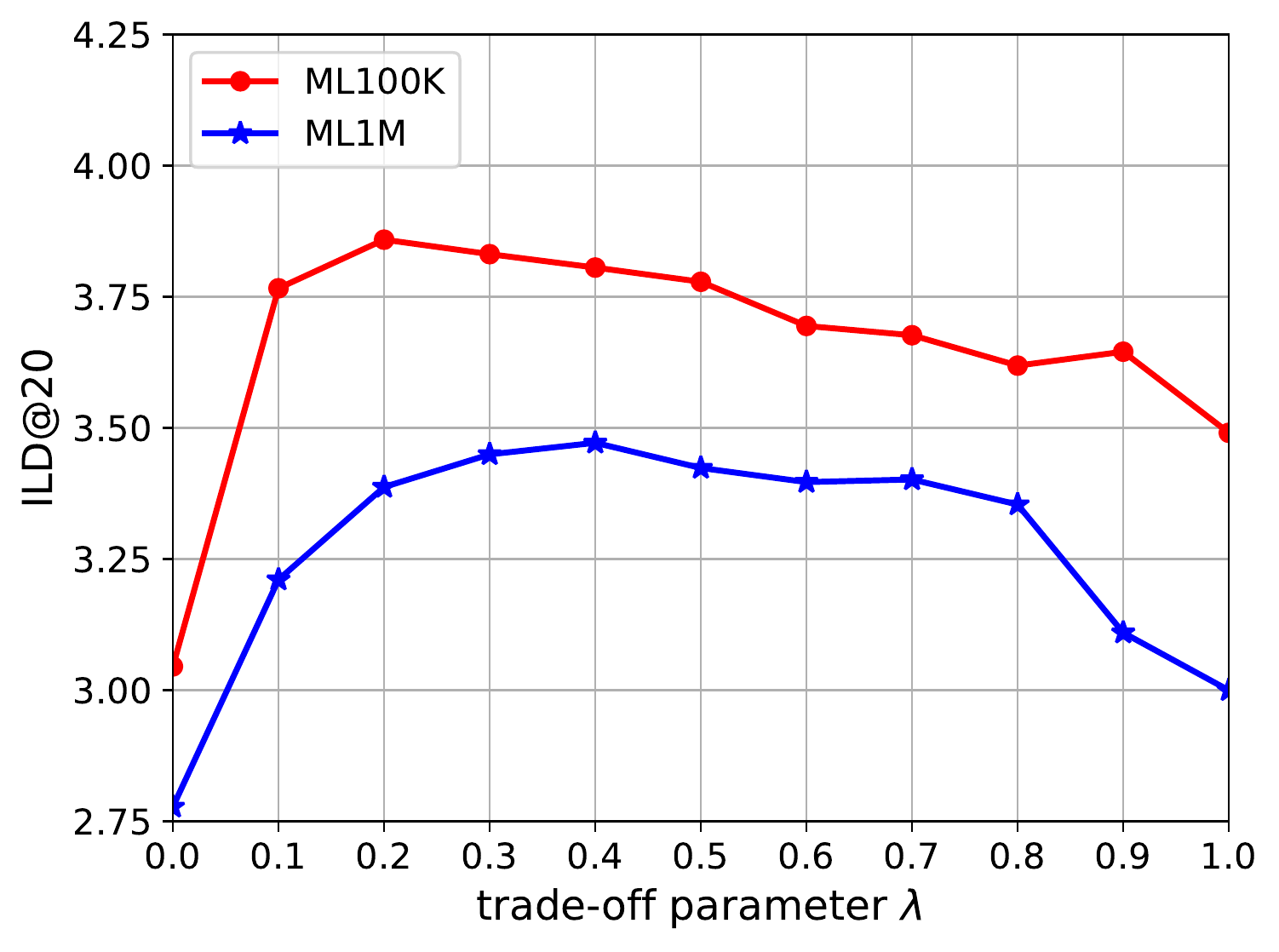}
                \caption{Performance in terms of ILD@20.}
                \label{ILD}
        \end{subfigure}
                \smallskip
                \caption{Performance of \ac{IDSR} on two datasets with the parameter $\lambda$ in Eq.~\eqref{equation11} changing from $0$ to $1$.}
\label{lambda} 
\end{figure*}

\ac{IDSR} shows a better performance with \ac{GRU} than with Transformer in terms of all metrics on both datasets.
We believe the reason is that Transformers rely on position embeddings to capture sequential information which are less effective than \ac{GRU}, especially when Transformers are not pre-trained on a large-scale dataset.
On the contrast, the recurrent nature of \ac{GRU} is especially designed for sequence modeling, which means it needs fewer data to capture the sequential information. 
This can be proved by the fact that the improvements of {DSR}$_{\mathit{GRU}}$ over {DSR}$_{\mathit{Trans}}$ on ML1M are smaller than that on ML100K. 
For example, the improvements in terms of Recall@20 are 10.94\% and 5.69\% on ML100K and ML1M, respectively.

\subsection{Influence of trade-off parameter $\lambda$}

To answer RQ4, we investigate the impact of $\lambda$ on \ac{IDSR} by ranging it from  $0$ to $1$ with a step size of $0.1$.
The results are shown in Fig.~\ref{lambda}.

We can see that the accuracy metrics, i.e., Recall@20 and MRR@20, show upward trends generally when $\lambda$ increases from $0$ to $1$. 
When $\lambda=0$, \ac{IDSR} shows the worst performance. 
However, a noticeable increase is observed when $\lambda$ changes from $0$ to $0.1$.
It is because that $\lambda=0$ means we only consider diversity without accuracy, thus the model cannot be trained well to recommend relevant items. 
\ac{IDSR} shows its best performance in terms of accuracy metrics with $\lambda$ at around 0.5 on ML100K, while around 0.8 on ML1M dataset.

Regarding recommendation diversity, \ac{IDSR} increases when $\lambda$ changes from 0 to 0.1 and then decreases from 0.4 (0.2) to 1 in terms of ILD@20 on both datasets. 
The best diversification performance of \ac{IDSR} appears with a small value of $\lambda$, i.e., 0.2 on ML100K and 0.4 on ML1M. 
When $\lambda$ is set to 1, ILD@20 of \ac{IDSR} drops significantly as it indicates that we do not consider diversification in our IDP loss.
To conclude, Fig.~\ref{lambda} demonstrates that our designed IDP loss can boost the performance of \ac{IDSR} when we take both of accuracy and diversification into consideration simultaneously.

\subsection{Case study}
\begin{figure}[!ht]
 \centering
   \includegraphics[clip,trim=0mm 0mm 0mm 0mm,width=0.3\textwidth]{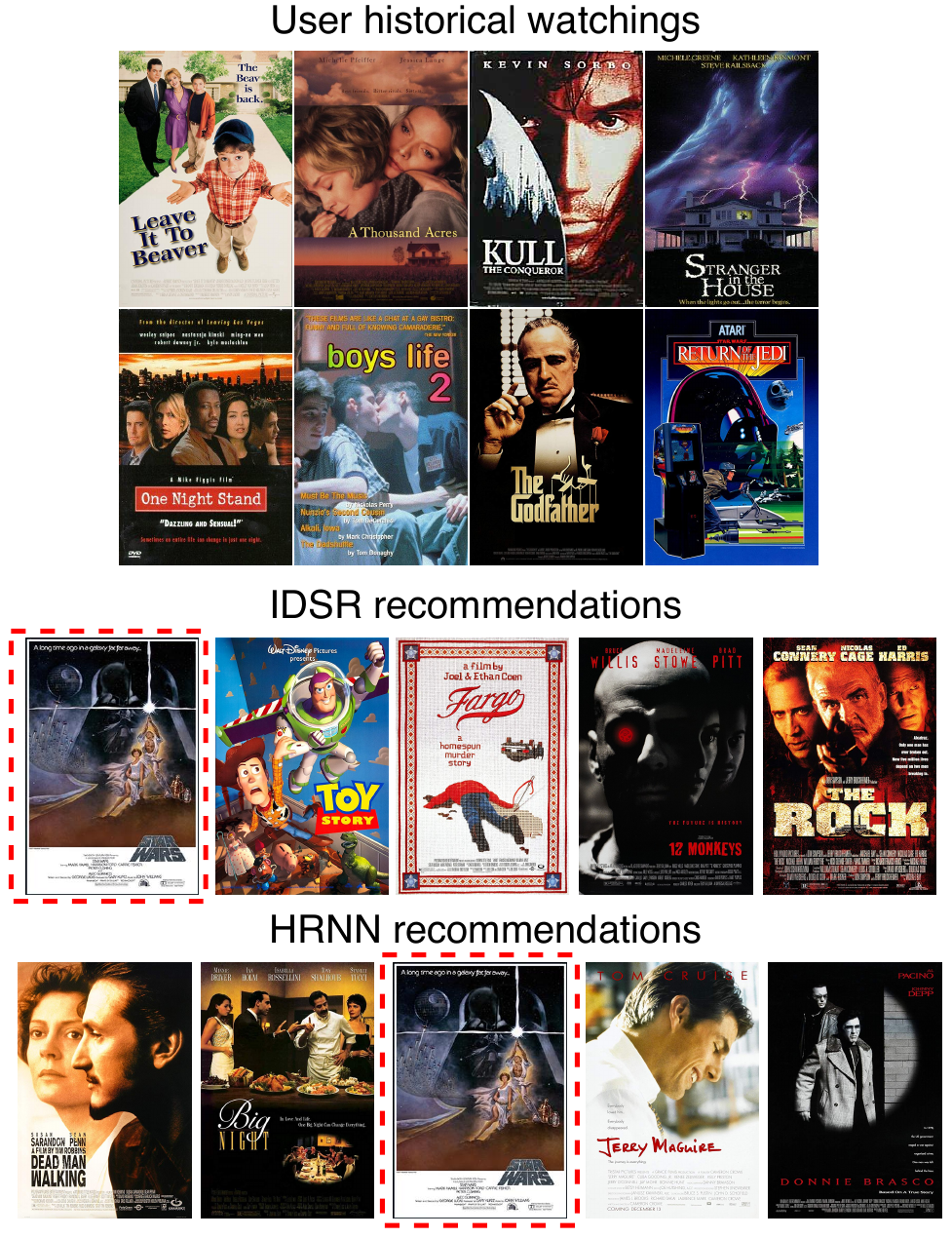}
   \caption{An example of recommendation results generated by \ac{IDSR} and HRNN.}
\label{case}
\end{figure}

To answer RQ5, we select an example from the test set of ML100K to illustrate the different recommendation results by \ac{IDSR} and HRNN in Fig.~\ref{case}. 

Fig.~\ref{case} shows 8 movies that the user watched recently and the top 5 recommendations generated by \ac{IDSR} and HRNN models, respectively. 
The ground truth item is marked with red box.
According to the user's historical watchings, we can tell that the user like Drama the most. 
But the user also shows interest in Comedy, Action, Thriller.
HRNN recommends four Drama movies which only takes care of the main intent of this user.
Differently, \ac{IDSR} accommodates all intents and diversifies the recommendation list with Drama, Comedy, Action, Adventure and Thriller movies.
Especially, \ac{IDSR} also recognizes the most important intent and rank a Drama movie at the top.
This confirms that \ac{IDSR} can not only mine users' implicit intents, but also generate a diversified recommendation list to cover those intents. 

\section{Conclusion and Future Work}
\label{conclusion}
In this paper, we propose \ac{IDSR} model to improve the diversification for \acp{SR}.
We devise the \ac{IIM} module to capture users' multiple intents and \ac{IDP} decoder to generate a diversified recommendation list covering those intents.
We also design an \ac{IDP} loss to supervise the model to consider accuracy and diversification simultaneously during training.
Our experimental results and in-depth analyses confirm the effectiveness of \ac{IDSR} on two datasets. 

As to future work, on the one hand, we plan to apply our model to other recommendation scenes, e.g., shared-account recommendations where the behaviors can be from multiple users with totally different intents. 
On the other hand, we hope to improve the recommendation accuracy by incorporating useful mechanisms from recent \ac{SR} models into \ac{IDSR}.

\newpage
\bibliographystyle{aaai}
\bibliography{bibliography}

\begin{thebibliography}{}

\bibitem[\protect\citeauthoryear{Abid \bgroup et al\mbox.\egroup
  }{2016}]{Abid2016}
Abid, A.; Hussain, N.; Abid, K.; Ahmad, F.; Farooq, M.~S.; Farooq, U.; Khan,
  S.~A.; Khan, Y.~D.; Naeem, M.~A.; and Sabir, N.
\newblock 2016.
\newblock A survey on search results diversification techniques.
\newblock {\em Neural Computing and Applications} 27(5):1207--1229.

\bibitem[\protect\citeauthoryear{Adomavicius and Tuzhilin}{2005}]{RS-CF2005}
Adomavicius, G., and Tuzhilin, A.
\newblock 2005.
\newblock Toward the next generation of recommender systems: A survey of the
  state-of-the-art and possible extensions.
\newblock {\em IEEE Transactions on Knowledge and Data Engineering.}
  17(6):734--749.

\bibitem[\protect\citeauthoryear{Agrawal \bgroup et al\mbox.\egroup
  }{2009}]{Agrawal:2009:DSR}
Agrawal, R.; Gollapudi, S.; Halverson, A.; and Ieong, S.
\newblock 2009.
\newblock Diversifying search results.
\newblock In {\em WSDM '09},  5--14.
\newblock New York, NY, USA: ACM.

\bibitem[\protect\citeauthoryear{Ashkan \bgroup et al\mbox.\egroup
  }{2015}]{Ashkan:2015:OGD}
Ashkan, A.; Kveton, B.; Berkovsky, S.; and Wen, Z.
\newblock 2015.
\newblock Optimal greedy diversity for recommendation.
\newblock In {\em IJCAI'15},  1742--1748.
\newblock AAAI Press.

\bibitem[\protect\citeauthoryear{Carbonell and Goldstein}{1998}]{MMR98}
Carbonell, J., and Goldstein, J.
\newblock 1998.
\newblock The use of mmr, diversity-based reranking for reordering documents
  and producing summaries.
\newblock In {\em SIGIR '98},  335--336.
\newblock New York, NY, USA: ACM.

\bibitem[\protect\citeauthoryear{Chen, Zhang, and Zhou}{2018}]{Chen:2018:FGM}
Chen, L.; Zhang, G.; and Zhou, H.
\newblock 2018.
\newblock Fast greedy map inference for determinantal point process to improve
  recommendation diversity.
\newblock In {\em NIPS'18},  5627--5638.
\newblock USA: Curran Associates Inc.

\bibitem[\protect\citeauthoryear{Cheng \bgroup et al\mbox.\egroup
  }{2017}]{Cheng:2017:LRA}
Cheng, P.; Wang, S.; Ma, J.; Sun, J.; and Xiong, H.
\newblock 2017.
\newblock Learning to recommend accurate and diverse items.
\newblock In {\em WWW '17},  183--192.

\bibitem[\protect\citeauthoryear{Glorot and Bengio}{2010}]{pmlr-v9-glorot10a}
Glorot, X., and Bengio, Y.
\newblock 2010.
\newblock Understanding the difficulty of training deep feedforward neural
  networks.
\newblock In {\em AI\&Statistics '10},  249--256.
\newblock Chia Laguna Resort, Sardinia, Italy: PMLR.

\bibitem[\protect\citeauthoryear{He and McAuley}{2016}]{Fusing2016}
He, R., and McAuley, J.
\newblock 2016.
\newblock Fusing similarity models with markov chains for sparse sequential
  recommendation.
\newblock In {\em International Conference on Data Mining},  191--200.
\newblock IEEE.

\bibitem[\protect\citeauthoryear{He \bgroup et al\mbox.\egroup
  }{2018}]{he2018nais}
He, X.; He, Z.; Song, J.; Liu, Z.; Jiang, Y.; and Chua, T.
\newblock 2018.
\newblock Nais: Neural attentive item similarity model for recommendation.
\newblock {\em IEEE Transactions on Knowledge and Data Engineering}
  30(12):2354--2366.

\bibitem[\protect\citeauthoryear{Hidasi \bgroup et al\mbox.\egroup
  }{2016a}]{2016session-based}
Hidasi, B.; Karatzoglou, A.; Baltrunas, L.; and Tikk, D.
\newblock 2016a.
\newblock Session-based recommendations with recurrent neural networks.
\newblock In {\em ICLR '16}.

\bibitem[\protect\citeauthoryear{Hidasi \bgroup et al\mbox.\egroup
  }{2016b}]{GRU2016}
Hidasi, B.; Karatzoglou, A.; Baltrunas, L.; and Tikk, D.
\newblock 2016b.
\newblock Session-based recommendations with recurrent neural networks.
\newblock In {\em ICLR'16},  1--10.

\bibitem[\protect\citeauthoryear{Kang and McAuley}{2018}]{self-attentive18}
Kang, W., and McAuley, J.~J.
\newblock 2018.
\newblock Self-attentive sequential recommendation.
\newblock In {\em ICDM '18},  197--206.

\bibitem[\protect\citeauthoryear{Kingma and Ba}{2014}]{Adam2014}
Kingma, D., and Ba, J.
\newblock 2014.
\newblock Adam: A method for stochastic optimization.
\newblock {\em arXiv preprint arXiv:1412.6980}.

\bibitem[\protect\citeauthoryear{Kulesza and Taskar}{2012}]{Kulesza:2012:DPP}
Kulesza, A., and Taskar, B.
\newblock 2012.
\newblock {\em Determinantal Point Processes for Machine Learning}.
\newblock Hanover, MA, USA: Now Publishers Inc.

\bibitem[\protect\citeauthoryear{Kunaver and Porl}{2017}]{diversity-survey}
Kunaver, M., and Porl, T.
\newblock 2017.
\newblock Diversity in recommender systems a survey.
\newblock {\em Know.-Based Syst.} 123(C):154--162.

\bibitem[\protect\citeauthoryear{Li \bgroup et al\mbox.\egroup
  }{2017}]{NARM2017}
Li, J.; Ren, P.; Chen, Z.; Ren, Z.; Lian, T.; and Ma, J.
\newblock 2017.
\newblock Neural attentive session-based recommendation.
\newblock In {\em CIKM '17},  1419--1428.
\newblock New York, NY, USA: ACM.

\bibitem[\protect\citeauthoryear{Liu \bgroup et al\mbox.\egroup
  }{2018}]{STAMP2018}
Liu, Q.; Zeng, Y.; Mokhosi, R.; and Zhang, H.
\newblock 2018.
\newblock Stamp: Short-term attention/memory priority model for session-based
  recommendation.
\newblock In {\em KDD '18},  1831--1839.
\newblock New York, NY, USA: ACM.

\bibitem[\protect\citeauthoryear{Quadrana \bgroup et al\mbox.\egroup
  }{2017}]{HRNN2017}
Quadrana, M.; Karatzoglou, A.; Hidasi, B.; and Cremonesi, P.
\newblock 2017.
\newblock Personalizing session-based recommendations with hierarchical
  recurrent neural networks.
\newblock In {\em RecSys '17},  130--137.
\newblock New York, NY, USA: ACM.

\bibitem[\protect\citeauthoryear{Quadrana, Cremonesi, and
  Jannach}{2018}]{Quadrana:2018:SRS:3236632.3190616}
Quadrana, M.; Cremonesi, P.; and Jannach, D.
\newblock 2018.
\newblock Sequence-aware recommender systems.
\newblock {\em ACM Computing Surveys} 51(4):66:1--66:36.

\bibitem[\protect\citeauthoryear{Rakkappan and
  Rajan}{2019}]{Rakkappan:2019:CSR:3308558.3313567}
Rakkappan, L., and Rajan, V.
\newblock 2019.
\newblock Context-aware sequential recommendations with stacked recurrent
  neural networks.
\newblock In {\em The Web Conference},  3172--3178.

\bibitem[\protect\citeauthoryear{Ren \bgroup et al\mbox.\egroup
  }{2019}]{ren-2019-repeatnet}
Ren, P.; Chen, Z.; Li, J.; Ren, Z.; Ma, J.; and de~Rijke, M.
\newblock 2019.
\newblock Repeatnet: A repeat aware neural recommendation machine for
  session-based recommendation.
\newblock In {\em AAAI '19}.
\newblock AAAI.

\bibitem[\protect\citeauthoryear{Rendle, Freudenthaler, and
  Schmidt-Thieme}{2010}]{FPMC2010}
Rendle, S.; Freudenthaler, C.; and Schmidt-Thieme, L.
\newblock 2010.
\newblock Factorizing personalized markov chains for next-basket
  recommendation.
\newblock In {\em WWW '10},  811--820.
\newblock New York, NY, USA: ACM.

\bibitem[\protect\citeauthoryear{Sha, Wu, and Niu}{2016}]{Sha:2016:FRR}
Sha, C.; Wu, X.; and Niu, J.
\newblock 2016.
\newblock A framework for recommending relevant and diverse items.
\newblock In {\em IJCAI'16},  3868--3874.
\newblock AAAI Press.

\bibitem[\protect\citeauthoryear{Sun \bgroup et al\mbox.\egroup
  }{2019}]{BERT2019}
Sun, F.; Liu, J.; Wu, J.; Pei, C.; Lin, X.; Ou, W.; and Jiang, P.
\newblock 2019.
\newblock Bert4rec: Sequential recommendation with bidirectional encoder
  representations from transformer.
\newblock {\em CoRR} abs/1904.06690.

\bibitem[\protect\citeauthoryear{Vaswani \bgroup et al\mbox.\egroup
  }{2017a}]{ATTention2017}
Vaswani, A.; Shazeer, N.; Parmar, N.; Uszkoreit, J.; Jones, L.; Gomez, A.~N.;
  Kaiser, L.; and Polosukhin, I.
\newblock 2017a.
\newblock Attention is all you need.
\newblock {\em CoRR} abs/1706.03762.

\bibitem[\protect\citeauthoryear{Vaswani \bgroup et al\mbox.\egroup
  }{2017b}]{NIPS2017_7181}
Vaswani, A.; Shazeer, N.; Parmar, N.; Uszkoreit, J.; Jones, L.; Gomez, Aidan
  Nand~Kaiser, L.~u.; and Polosukhin, I.
\newblock 2017b.
\newblock Attention is all you need.
\newblock In {\em Advances in Neural Information Processing Systems 30},
  5998--6008.
\newblock Curran Associates, Inc.

\bibitem[\protect\citeauthoryear{Wu \bgroup et al\mbox.\egroup
  }{2019}]{diversity-advances}
Wu, Q.; Liu, Y.; Miao, C.; Zhao, Y.; Guan, L.; and Tang, H.
\newblock 2019.
\newblock Recent advances in diversified recommendation.
\newblock {\em CoRR} abs/1905.06589.

\bibitem[\protect\citeauthoryear{Xu \bgroup et al\mbox.\egroup
  }{2019}]{Xu:2019:RCN}
Xu, C.; Zhao, P.; Liu, Y.; Xu, J.; S.Sheng, V.~S.; Cui, Z.; Zhou, X.; and
  Xiong, H.
\newblock 2019.
\newblock Recurrent convolutional neural network for sequential recommendation.
\newblock In {\em WWW '19},  3398--3404.
\newblock New York, NY, USA: ACM.

\bibitem[\protect\citeauthoryear{Zhang and Hurley}{2008}]{Zhang:2008:AMI}
Zhang, M., and Hurley, N.
\newblock 2008.
\newblock Avoiding monotony: Improving the diversity of recommendation lists.
\newblock In {\em RecSys '08},  123--130.
\newblock New York, NY, USA: ACM.

\bibitem[\protect\citeauthoryear{Zhang, Yao, and Sun}{2017}]{RSsurvey2017}
Zhang, S.; Yao, L.; and Sun, A.
\newblock 2017.
\newblock Deep learning based recommender system: {A} survey and new
  perspectives.
\newblock {\em arXiv preprint arXiv:1707.07435}.

\bibitem[\protect\citeauthoryear{Zimdars, Chickering, and
  Meek}{2001}]{Zimdars:2001}
Zimdars, A.; Chickering, D.~M.; and Meek, C.
\newblock 2001.
\newblock Using temporal data for making recommendations.
\newblock In {\em UAI '01},  580--588.
\newblock San Francisco, CA, USA: Morgan Kaufmann Publishers Inc.

\end{thebibliography}

\end{document}